\documentclass[conference]{IEEEtran}

\usepackage{amsmath,amssymb,graphicx,subcaption}

\usepackage{booktabs,multirow,tabularx,makecell}
\usepackage{siunitx}   % neat numbers/alignment
\newcolumntype{Y}{>{\raggedright\arraybackslash}X}   % ragged-right text col
\newcommand{\tid}[1]{\textbf{T#1}}                   % bold target IDs
\newcommand{\phsum}[1]{\emph{#1}}       % phonetic short note
\usepackage{pgfplots}
\usepackage{pgfplotstable}
\usepackage{booktabs}
\usepackage{caption}
\usepackage{subcaption}
\usepackage{url}
\usepackage{balance}
\usepackage{phonetic}
\usepackage{pgfplotstable}
\usepackage{booktabs}
\usepackage{array} % optional, nice column tweaks

\pgfplotsset{compat=1.18}

\title{Impact of Phonetics on Speaker Identity in Adversarial Voice Attack}

\author{
\IEEEauthorblockN{Daniyal Kabir Dar, Qiben Yan, Li Xiao, Arun Ross}
\IEEEauthorblockA{Department of Computer Science and Engineering, Michigan State University, USA}
\IEEEauthorblockA{\{dardaniy, qyan, lxiao, rossarun\}@msu.edu}
}
\begin{document}

\maketitle

\begin{abstract}
Adversarial perturbations in speech pose a serious threat to automatic speech recognition (ASR) and speaker verification by introducing subtle waveform modifications that remain imperceptible to humans but can significantly alter system outputs. While targeted attacks on end-to-end ASR models have been widely studied, the phonetic basis of these perturbations and their effect on speaker identity remain underexplored. In this work, we analyze adversarial audio at the phonetic level and show that perturbations are associated with systematic phonetic tendencies, such as vowel centralization and consonant substitutions.
These distortions not only mislead transcription but also degrade phonetic cues critical for speaker verification, leading to \emph{identity drift}. Using the DeepSpeech ASR model as our target, we generate targeted adversarial examples and evaluate their impact on speaker identity embeddings across genuine and impostor samples. Results across 16 phonetically diverse target phrases demonstrate that adversarial audio induces both transcription errors and identity drift, highlighting the need for phonetic-aware defenses to ensure the robustness of ASR and speaker recognition systems.
\end{abstract}

\begin{IEEEkeywords}
Adversarial attack, phonetics, identity drift, speech transcription
\end{IEEEkeywords}

\section{Introduction}
Automatic speech recognition (ASR) has achieved remarkable progress with the advent of deep learning, enabling deployment in applications ranging from voice assistants and transcription services to security-critical domains such as speaker authentication and forensic analysis. Despite these advances, recent research has demonstrated that ASR systems are highly vulnerable to adversarial perturbations—imperceptible modifications to the input waveform that can cause dramatic transcription errors \cite{carlini2018audio}. Such vulnerabilities raise fundamental concerns about the reliability and security of speech technologies.  

\begin{figure}[!t] % note the * for two-column placement
\centering
\includegraphics[width=0.5\textwidth]{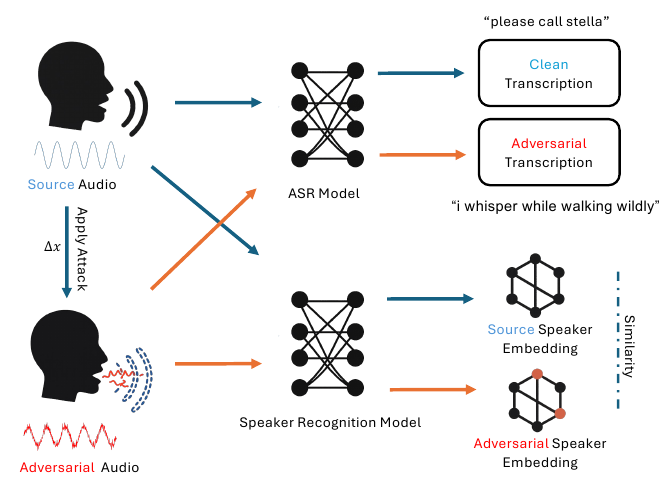}
\caption{Overview of the adversarial audio attack framework. 
A perturbation $\Delta x$ is applied to source audio to produce adversarial audio. 
The Automatic Speech Recognition (ASR) model yields adversarial transcriptions, while the Speaker Recognition model produces embeddings that drift from the source identity, degrading biometric similarity.}
\label{fig:system_overview}
\end{figure}

While prior work has primarily characterized adversarial audio in terms of transcription metrics such as word error rate (WER) and character error rate (CER), comparatively little attention has been paid to the phonetic dimension of these perturbations. Phonetic units serve as the bridge between acoustic signals and linguistic representation; perturbations that alter phoneme realizations can not only change transcribed text but also modify salient cues used to characterize speaker identity. This dual effect suggests that adversarial perturbations may induce \emph{identity drift}: a shift in the acoustic–phonetic footprint of a speaker that degrades the stability of downstream biometric speaker recognition. While attack success is defined by forced transcription, identity drift is quantified post-attack via changes in speaker embedding similarity.

In this paper, we take a phonetic perspective on adversarial attacks against ASR. Using DeepSpeech \cite{Hannun2014DeepSS} as our target model, we generate targeted adversarial examples and analyze their impact at two levels: phoneme-level confusions, and speaker identity embeddings. Our results show that adversarial perturbations are often associated with systematic phonetic confusions, such as vowel centralization and consonant substitutions, which propagate into altered speaker embeddings. This not only produces misleading transcriptions but also erodes the consistency of speaker identity, demonstrating that adversarial audio attacks compromise ASR and speaker recognition simultaneously.

The contributions of this work are threefold: 
(i) we provide a phonetic-level analysis of adversarial audio, revealing how perturbations exploit systematic phoneme confusions; 
(ii) we utilize the concept of \emph{identity drift}, characterizing how adversarial perturbations alter phonetic cues underlying speaker identity; and 
(iii) we conduct extensive experiments on DeepSpeech and evaluate speaker recognition metrics, demonstrating the dual impact of adversarial perturbations.

\section{Related Work}
We evaluate the effect of adversarial perturbations on speaker identity using two state-of-the-art identity embedding models: ECAPA-TDNN \cite{desplanques2020ecapa} and a ResNet-based speaker verification backbone \cite{heigold2016resnet, chung2020voxceleb}. These findings build on a large body of work showing that both ASR and speaker verification (SV) systems are vulnerable to adversarial manipulation. Hidden and practical command injection attacks such as CommanderSong \cite{yuan2018commandersong}, DolphinAttack \cite{zhang2017dolphin}, Hidden Voice Commands \cite{carlini2016hidden}, CommanderGabble \cite{zhang2021commandergabble}, and recent phoneme-level manipulations like PhantomSound \cite{guo2023phantomsound} and SMACK \cite{yu2023smack} highlight real-world feasibility. Targeted adversarial examples against ASR were first demonstrated by Carlini and Wagner \cite{carlini2018audio}, with subsequent advances leveraging psychoacoustic models, masking, and over-the-air robustness \cite{schonherr2018adversarial, qin2019imperceptible, yakura2019robust}. More recent attack strategies include semantically meaningful perturbations \cite{yu2023smack}, query-efficient black-box methods \cite{guo2023phantomsound}, and jailbreak-style attacks on audio-language models \cite{kang2024advwave}. Localized and structured adversarial attacks have also been explored in the spectrogram domain, such as SpecPatch \cite{guo2022specpatch}, which introduces a human-in-the-loop framework for crafting adversarial spectrogram patches against ASR systems.
Speaker verification systems face similar threats: early works demonstrated targeted attacks against i-vector and DNN-based SV \cite{li2020practical, xie2020realtime, chen2021who}, with extensions to universal perturbations \cite{zhang2021commandergabble}, robust real-time variants \cite{xie2020realtime}, and stress-testing embedding models \cite{kreuk2018fooling, li2020adversarial, todisco2024malacopula}. Attacks exploiting voice conversion and spoofing further degrade reliability \cite{cai2023identifying, blue2022you}, while studies such as Zhang et al. \cite{zhang2023impact} demonstrate how silence and phonetic boundaries influence spoofing detection. On the defense side, Wu et al. \cite{wu2021improving} improved robustness via self-supervised pretraining, Chen et al. \cite{chen2024neuralcodec} leveraged codec bottlenecks for adversarial detection, and Kühne et al. \cite{kuhne2025detecting} introduced diffusion models for ASR defenses. Complementary work includes DeepDetection for phishing prevention \cite{app122111109}, vocal tract reconstruction for deepfake detection \cite{blue2022you}, and codec- or vocoder-based anti-spoofing pipelines \cite{chen2024neuralcodec, zhang2023impact}. Despite this breadth, most approaches remain agnostic to the phonetic underpinnings of perturbations. Our work addresses this gap by explicitly connecting phoneme-level distortions—such as fricative instability, stop consonant confusion, and vowel centralization—to embedding drift, utilizing the concept of \emph{identity drift} to unify transcription and speaker verification vulnerabilities.

\section{Methodology}

\subsection{Problem Formulation}
Let $x \in \mathbb{R}^T$ denote an input audio waveform of length $T$. Our goal is to construct an adversarial example $x' = x + \delta$ such that $x'$ is perceptually indistinguishable from $x$ ($\|\delta\|$ is small under a distortion metric), yet simultaneously causes two effects: (i) the ASR system transcribes $x'$ as a chosen target phrase $y_t$, and (ii) the speaker identity associated with $x$ drifts in the embedding space of a speaker recognition system. This second phenomenon is known as \emph{identity drift}. Formally, an attack succeeds if
\[
C(x') = y_t \quad \text{and} \quad d(f(x), f(x')) \gg 0,
\]
where $C(\cdot)$ denotes the ASR transcription function, $f(\cdot)$ the speaker embedding extractor, and $d(\cdot,\cdot)$ a similarity measure.

\subsection{Adversarial Attack on DeepSpeech}
We adopt a white-box targeted attack on the pretrained DeepSpeech end-to-end ASR model. DeepSpeech is trained using the Connectionist Temporal Classification (CTC) loss, which aligns variable-length audio frames with target transcripts. Following Carlini \& Wagner \cite{carlini2018audio}, we optimize perturbations $\delta$ using an iterative gradient-based method:
\[
\min_{\delta} \; \|\delta\|_2^2 + c \cdot \text{CTC-Loss}(x+\delta, y_t),
\]
subject to $x+\delta \in [-M, M]$, where $M$ is the maximum representable audio amplitude. Distortion is quantified in terms of signal-to-noise ratio (SNR, dB), ensuring perturbations remain imperceptible to human listeners. We generate targeted adversarial examples on individual utterances to evaluate transcription errors and their effect on speaker identity embeddings.

\subsection{Target Set and Phonetic Coverage}

To systematically evaluate adversarial perturbations, we curated a set of 16 target transcriptions (T1–T16) that balance short commands, medium-length control phrases, and long pangram-style sentences. This design ensures coverage of a wide range of phonetic phenomena, including different vowel qualities, consonant manners (plosives, fricatives, affricates, nasals, approximants), stress patterns, and prosodic structures. The inclusion of pangrams and alliterative sentences allows us to probe how phonetic density and articulatory demands affect both transcription robustness and speaker identity verification. This design allows us to explicitly connect attack outcomes with phonetic structure.

For clarity, Table~\ref{tab:targets-metrics} reports results for ten representative targets: T1, T2, T3, T4, T5 , T6, T9, T10, T12 and T14. The remaining targets, not shown in the table, include:  
T7 (\emph{two tall teachers talk to Tim}),  
T8 (\emph{I whisper while walking wildly}),    
T11 (\emph{a mad boxer shot a quick gloved jab to the jaw of his dizzy opponent}),  
T13 (\emph{twelve jolly grizzlies briskly danced over waxy benches while a flighty kitten kept humming jazz tunes in the background}),  
T15 (\emph{while whispering winds wander westward jittery jackals jiggled jellies above velvet jars beyond flickering bonfires in a frozen jungle}), and  
T16 (\emph{kindly expedite bizarre frozen jumpsuits for victors whirlwind gala to maximize xenon emissions before daybreak}). Together, these 16 targets cover fricative clusters, stop alliteration, vowel-rich phrases, and highly complex pangrams. For each case, we report the number of samples, genuine and impostor sample counts, True Match Rate (TMR) at 0.1\% False Match Rate (FMR), and the discriminability index $d'$ as a measure of the overlap between genuine and impostor score distributions, enabling a direct connection between phonetic composition and observed identity drift. The phonetic profiles highlight which classes of sounds dominate each phrase (e.g., fricative clusters, voiced vs.\ unvoiced stops, vowel richness), enabling a direct connection between phonetic composition and observed identity drift. This unified presentation provides both a linguistic rationale and an empirical grounding for our analysis.

\subsection{Speaker Identity Evaluation}
We evaluate the effect of adversarial perturbations on speaker identity using two state-of-the-art embedding models: ECAPA-TDNN \cite{desplanques2020ecapa} and a ResNet-based speaker verification backbone \cite{heigold2016resnet, chung2020voxceleb}. Given an utterance $x$, each model outputs a fixed-dimensional embedding $f(x) \in \mathbb{R}^d$. We assess the similarity between embeddings using cosine distance. Two types of samples are considered: \emph{genuine samples} (same speaker: $f(x)$ vs. $f(x')$) and \emph{impostor samples} (different speakers). 
To quantify the separation between the genuine and impostor score distributions, we compute the discriminability index $d'$\cite{macmillan2004detection}, defined as:

\[
d' = \frac{\mu_{\text{gen}} - \mu_{\text{imp}}}{\sqrt{\tfrac{1}{2}(\sigma_{\text{gen}}^2 + \sigma_{\text{imp}}^2)}},
\]
where $\mu_{\text{gen}}$ and $\mu_{\text{imp}}$ denote the means of the genuine and impostor score distributions, respectively, and $\sigma_{\text{gen}}^2$ and $\sigma_{\text{imp}}^2$ are their variances. A significant drop in $d'$ indicates that adversarial perturbations shift embeddings closer to impostor space, thereby inducing identity drift. In addition, we report the TMR at 0.1\% FMR. Using both ECAPA-TDNN and ResNet allows us to validate that observed identity drift is not model-specific but reflects vulnerabilities of embedding-based speaker recognition systems.

% In addition, we report the true match rate at 0.1\% false match rate (TMR@0.1\%FMR). Using both ECAPA-TDNN and ResNet allows us to validate that observed identity drift is not model-specific but reflects vulnerabilities of embedding-based speaker recognition systems.

\begin{table*}[htbp]
\centering
\caption{Representative 10 targets with short phonetic profiles and identity metrics per model:
TMR measured at FMR $=0.1\%$ and $d'$ from genuine vs.\ impostor score distributions.}
\label{tab:targets-metrics}
\scriptsize
\setlength{\tabcolsep}{5pt}
\begin{tabularx}{\textwidth}{l Y Y l S[table-format=5.0] S[table-format=5.0] S[table-format=5.0]
                             l S[table-format=1.4] S[table-format=1.2]}
\toprule
\textbf{ID} & \textbf{Target text} & \textbf{Phonetic short note} & \textbf{Model}
& {\#Samples} & {\#Genuine} & {\#Impostor} & \textbf{TMR@0.1\%FMR} & {$d'$} \\
\midrule
\tid{1}  & yes & \phsum{mono-syll.; 1:2 V:C; glide+fric stop} 
         & ECAPA    & 11881 & 109 & 11772 & 1.0000 & 9.68 \\
         &     &                                              & ResNet50 & 11881 & 109 & 11772 & 1.0000 & 9.43 \\
\midrule
\tid{2}  & open the door & \phsum{4 syll.; 4:6 V:C; dental fric.\ + stops} 
         & ECAPA    & 11881 & 109 & 11772 & 1.0000 & 9.11 \\
         &               &                                                   & ResNet50 & 11881 & 109 & 11772 & 1.0000 & 8.89 \\
\midrule
\tid{3}  & call emergency services & \phsum{fricative-rich; 8 syll.; 8:16 V:C} 
         & ECAPA    & 11881 & 109 & 11772 & 0.9725 & 5.59 \\
         &                          &                                             & ResNet50 & 11881 & 109 & 11772 & 0.9817 & 6.22 \\
\midrule
\tid{4}  & the quick brown fox jumped over the lazy dog & \phsum{pangram; 11 syll.; broad coverage} 
         & ECAPA    & 11881 & 109 & 11772 & 0.9083 & 4.80 \\
         &                         &                                                       & ResNet50 & 11881 & 109 & 11772 & 0.9817 & 5.14 \\
\midrule
\tid{5}  & shhh she sees the sea fish & \phsum{fricative-rich: /sh, s, z/; 6 syll.}
 
         & ECAPA    & 11881 & 109 & 11772 & 0.9908 & 7.46 \\
         &                             &                                               & ResNet50 & 11881 & 109 & 11772 & 1.0000 & 7.74 \\
\midrule
\tid{6}  & do go big bag dig & \phsum{voiced stops chain; minimal vowels; 5 syll.} 
         & ECAPA    & 11881 & 109 & 11772 & 1.0000 & 8.44 \\
         &                    &                                                     & ResNet50 & 11881 & 109 & 11772 & 1.0000 & 8.35 \\
\midrule
\tid{9}  & pack my box with five dozen liquor jugs & \phsum{pangram; many consonant clusters} 
         & ECAPA & 9025 & 95 & 8930 & 0.8632 & 4.63 \\
         &                                         &                                    & ResNet50 & 9025 & 95 & 8930 & 0.9474 & 5.02 \\
\midrule
\tid{10} & glib jocks quiz nymph to vex dwarf & \phsum{pangram; high fricative/affricate load} 
         & ECAPA & 9025 & 95 & 8930 & 0.8421 & 4.75 \\
         &                                      &                                                   & ResNet50 & 9025 & 95 & 8930 & 0.9579 & 5.34 \\
\midrule
\tid{12} & just before twilight the wizard quickly jabbed five boxes of hazy quartz to vex a plump knight’s jovial frog 
         & \phsum{very long pangram; many clusters; vowel centralization} 
         & ECAPA & 6561 & 81 & 6480 & 0.4444 & 3.07 \\
         &      &                                                                                     & ResNet50 & 6561 & 81 & 6480 & 0.7160 & 3.63 \\
\midrule
\tid{14} & quantum driven flux engines jam beneath zigzagging vortex panels as cryptic bioforms whisper behind polymorphic glass domes 
         & \phsum{dense consonant clusters; many fricatives/affricates} 
         & ECAPA & 2209 & 47 & 2162 & 0.6809 & 3.10 \\
         &      &                                                                                     & ResNet50 & 2209 & 47 & 2162 & 0.7447 & 3.78 \\
\bottomrule
\end{tabularx}
\end{table*}

\begin{figure}[t]
    \centering
    \includegraphics[width=\linewidth]{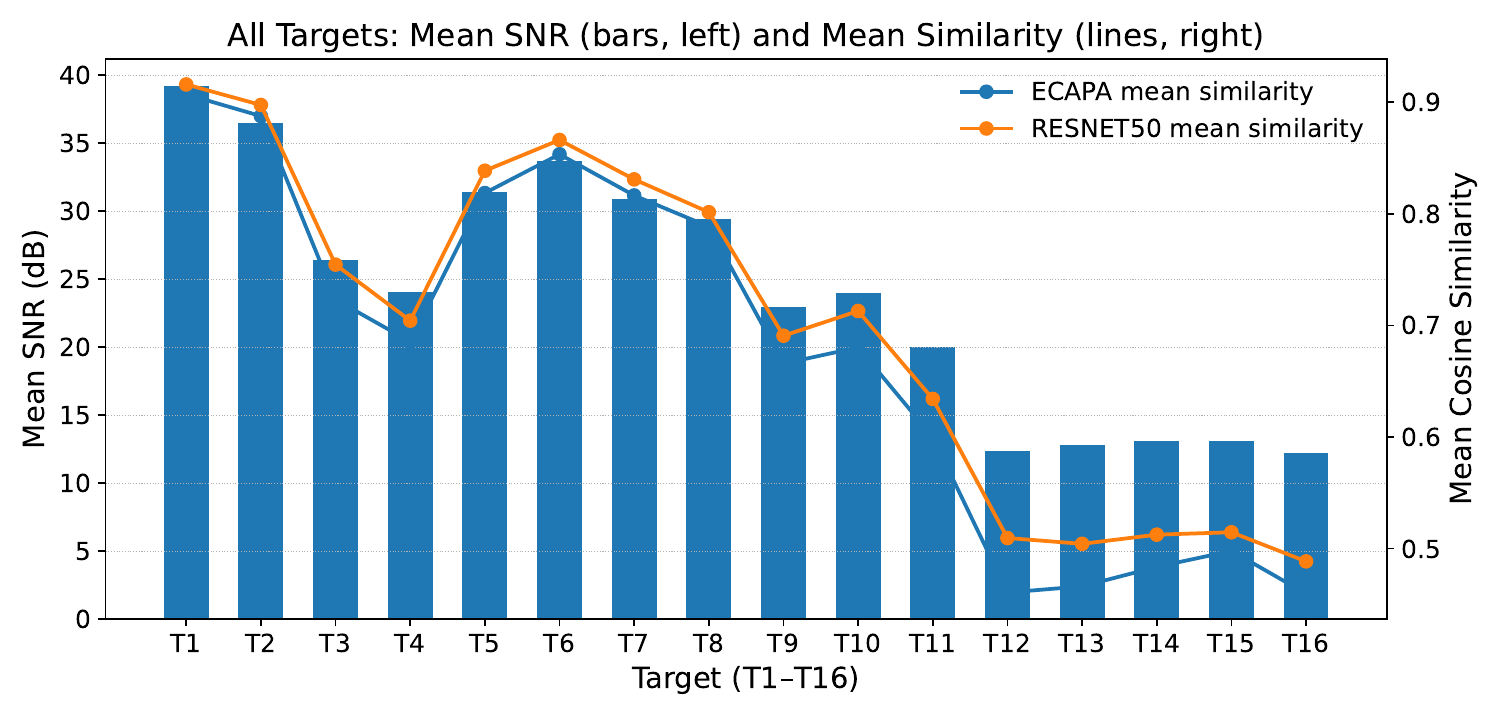}
    \caption{Mean SNR (bars, left axis) and mean cosine similarity (lines, right axis) 
    across all 16 target transcriptions (T1–T16). 
    Similarity trends are shown for both ECAPA and ResNet50 embeddings.}
    \label{fig:snr_similarity_all_targets}
\end{figure}

% --- Inline data (keeps things self-contained). You can move this to a CSV if you prefer. ---

% % --- Compact mapping table: Ti -> full target text (for reference in the paper) ---
% \begin{table}[t]
% \centering
% \pgfplotstabletypeset[
%   col sep=comma,
%   string type,
%   columns/T/.style={string type, column name={T}},
%   columns/Target/.style={string type, column name={Target}},
%   columns/ECAPA/.style={column name={ECAPA $d'$}},
%   columns/RESNET50/.style={column name={RESNET50 $d'$}},
%   columns={T,Target,ECAPA,RESNET50},
%   every head row/.style={before row=\toprule,after row=\midrule},
%   every last row/.style={after row=\bottomrule},
%   begin table=\small,]{dprime_targets.csv}
% \caption{Target texts and d-prime values per model (values used in Fig.~\ref{fig:dprime_grouped}).}
% \label{tab:dprime_targets}
% \end{table}

\section{Experiments and Results}

\noindent\textbf{Experimental Setup.} 
We conduct experiments on the VCTK corpus \cite{yamagishi2019vctk}, which contains recordings from 109 native English speakers with diverse accents. This dataset provides the multi-speaker coverage necessary to evaluate both adversarial attack success and speaker identity drift. We generate targeted adversarial examples on 16 transcriptions (T1--T16), including short commands (e.g., ``yes'', ``open the door''), medium-length phrases, and pangram-style sentences designed to maximize phonetic diversity. All attacks are implemented in a white-box setting against DeepSpeech, using iterative optimization until convergence on GPUs. For each of the 109 speaker IDs, we select a clean source utterance and attempt to drive that utterance to each of the 16 target transcriptions, yielding a maximum of $109\times16$ source--target pairs per model.
Having established the experimental protocol, we next examine the effectiveness of these attacks in terms of speaker identity verification.  
% --- Figure: grouped bars ECAPA vs RESNET50 across T1..T16 ---
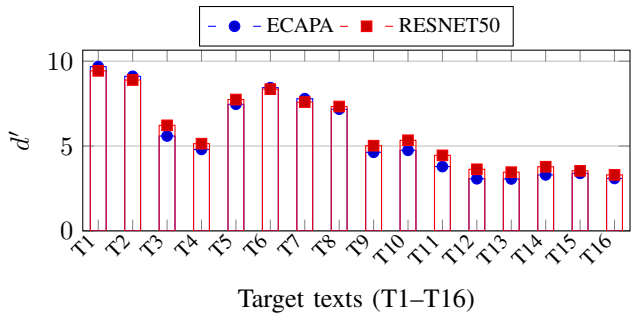
\begin{figure}[t]
  \centering
  \pgfplotstableread[col sep=comma]{dprime_targets.csv}\dprimetable
  \begin{tikzpicture}
    \begin{axis}[
      ymajorgrids,
      width=\linewidth,
      height=0.45\linewidth,
      ymin=0,
      ylabel={$d'$},
      xlabel={Target texts (T1--T16)},
      symbolic x coords={T1,T2,T3,T4,T5,T6,T7,T8,T9,T10,T11,T12,T13,T14,T15,T16},
      xtick=data,
      xticklabel style={font=\footnotesize, rotate=45, anchor=east},
      legend style={font=\footnotesize, at={(0.5,1.02)}, anchor=south, legend columns=2},
      bar width=6pt,
      enlarge x limits=0.03,
    ]
      \addplot+[ybar] table[x=T,y=ECAPA]{\dprimetable};
      \addplot+[ybar] table[x=T,y=RESNET50]{\dprimetable};
      \legend{ECAPA,RESNET50}
    \end{axis}
  \end{tikzpicture}
  \vspace{0.5em}
  \caption{$d'$ across target texts (T1--T16) for ECAPA vs.\ RESNET50.}
  \label{fig:dprime_grouped}
\end{figure}

\noindent\textbf{Speaker Identity Verification.} 
Short commands such as T1 and T2 preserved identity with TMR values of 100\%. Moderate-complexity phrases (T3, T5, T10) exhibited measurable drift, with TMR falling into the 85--99\% range. The longest utterances (T12--T16) collapsed to 44--75\% TMR, indicating severe overlap between genuine and impostor scores. This trend is mirrored in Fig.~\ref{fig:dprime_grouped}, where the discriminability index $d'$ drops from $\sim$9 for simple commands to $\sim$3 for the longest pangrams.  

\noindent\textbf{Phonetic and Length Correlations.} 
Phonetic composition strongly influenced robustness (Table~\ref{tab:targets-metrics}). Fricatives and affricates (T5, T10, T14) were disproportionately unstable, while voiced/unvoiced stop contrasts (T6, T7) blurred under perturbation. Vowel-rich phrases (T2, T8) remained comparatively robust. Utterance length further amplified drift: longer phrases consistently showed sharper TMR degradation and lower $d'$ values (e.g., T12 and T16).  

\noindent\textbf{Noise Level (SNR).} 
Perturbation strength also correlated inversely with verification performance. Figure~\ref{fig:snr_similarity_all_targets} summarizes the mean SNR and corresponding similarity values across all 16 target transcriptions. The bar plots show the average perturbation strength (SNR) per target, while the lines capture identity drift trends measured via cosine similarity under ECAPA and ResNet50 embeddings. At high SNR ($\sim$40 dB) perturbations remained imperceptible yet already degraded TMR for complex targets. At moderate SNR ($\sim$30 dB), drift worsened, especially in fricative-heavy phrases. At low SNR ($\sim$15 dB), identity drift became catastrophic, with TMR values below 53\% for T12 and T16.  

\noindent\textbf{Cross-Model Consistency.} 
Both ECAPA-TDNN and ResNet exhibited similar trends across the targets (Table~\ref{tab:targets-metrics}, Fig.~\ref{fig:dprime_grouped}), confirming that the observed vulnerabilities stem from phonetic and length structure rather than model-specific idiosyncrasies.

\section{Main Findings}
Our experiments reveal several consistent patterns. Targeted attacks not only succeed in forcing adversarial transcriptions but also induce measurable \emph{identity drift} in speaker embeddings. Phonetic structure plays a critical role: fricatives, affricates, and voiced/unvoiced stop contrasts are disproportionately unstable, whereas vowel-rich phrases remain more robust. Utterance length further amplifies drift, with TMR dropping below 50\% for extended pangrams. Perturbation strength also matters—identity drift is observed even at high SNR levels where perturbations are imperceptible, and drift severity increases as SNR decreases. Finally, both ECAPA and ResNet exhibit similar trends, indicating that identity drift is not model-specific but a general property of embedding-based speaker verification.

\section{Conclusion}
We present a phonetic perspective on adversarial attacks in the audio domain, demonstrating that perturbations not only mislead automatic speech recognition but also distort speaker embeddings in ways that compromise biometric verification. By systematically evaluating 16 phonetically diverse target phrases, we utilize the concept of \emph{identity drift}, showing that adversarial audio shifts speaker embeddings toward impostor space, significantly degrading the separability of genuine and impostor samples and that identity drift is strongly correlated with phonetic complexity and utterance length: short vowel-rich commands remained robust, while fricative-heavy, cluster-rich pangrams exhibited severe degradation in verification performance. These findings underscore that adversarial audio is not a uniform phenomenon but one shaped by linguistic structure and acoustic-phonetic cues.  

The importance of this work lies in revealing that adversarial attacks threaten the integrity of speech systems on multiple levels—semantics and identity—while remaining imperceptible to human listeners. To establish a foundation for this field, our study focused exclusively on \emph{white-box attacks} in controlled experimental settings. While this provides a clear view of underlying vulnerabilities, it also raises questions about real-world applicability, as over-the-air effects, environmental distortions, and black-box threat models remain unaddressed. Exploring these factors represents an important direction for future work.  

Future research should, therefore, extend beyond controlled conditions to investigate phonetic-aware defenses, robustness training that explicitly accounts for phoneme-level vulnerabilities, and detection methods sensitive to the subtle score distribution shifts we identified. More broadly, adversarial attacks on audio  warrant deeper study, particularly their interactions with linguistic variability, prosody, and channel effects. By bridging adversarial machine learning with phonetics, we take a step toward more secure and linguistically informed speech technologies.  
For reproducibility and extended visualizations, we provide additional figures in our GitHub repository\footnote{\url{https://daniyalkabir.github.io/icassp-2026-results/}}.

\section*{Acknowledgment}
This work was supported in part by the U.S. National Science Foundation grant CNS-2310207.

\balance
\bibliographystyle{IEEEtran}
\bibliography{references}

\end{document}